\documentclass[journal,12pt,draftclsnofoot,onecolumn]{IEEEtran}
\usepackage[utf8]{inputenc} 
\usepackage[T1]{fontenc}    
\usepackage{hyperref}       
\usepackage{url}            
\usepackage{booktabs}       
\usepackage{amsfonts}       
\usepackage{nicefrac}       
\usepackage{microtype}      
\usepackage{algorithm}
\usepackage{algorithmic}
\usepackage{xcolor}
\usepackage{cite}

\usepackage{microtype}
\usepackage{graphicx}
\usepackage{subfigure}
\usepackage{booktabs} 
\usepackage{amsmath}
\usepackage{amssymb}
\usepackage{amsthm}
\usepackage{bbm}

\newtheorem{theorem}{Theorem}

\newtheorem{proposition}[theorem]{Proposition}
\newtheorem{lemma}[theorem]{Lemma}

\DeclareMathOperator*{\argmax}{argmax}

\newcommand{\review}[1]{\textcolor{black}{#1}}

\ifCLASSINFOpdf
\else
\fi

\begin{document}
%
\title{Reinforcement Learning for Efficient and Tuning-Free Link Adaptation}
%
%
%

\author{Vidit~Saxena, Hugo~Tullberg, and~Joakim~Jald\'{e}n
\thanks{Vidit Saxena is with the Division of Information Science and Engineering, KTH Royal Institute of Technology, Stockholm, and with Ericsson Research, Stockholm. Dr. Hugo Tullberg is with Ericsson Research, Stockholm. Prof. Joakim Jald\'{e}n is with the Division of Information Science and Engineering, KTH Royal Institute of Technology, Stockholm. This work was partially supported by the Wallenberg Artificial Intelligence, Autonomous Systems and Software Program (WASP) funded by Knut and Alice Wallenberg Foundation, and by the European Research Council project AGNOSTIC (742648). This work has been submitted to the IEEE for possible publication. Copyright may be transferred without notice, after which this version may no longer be accessible.}}

\maketitle

\begin{abstract}
    \review{Wireless links adapt the data transmission parameters to the dynamic channel state -- this is called \emph{link adaptation}. Classical link adaptation relies on tuning parameters that are challenging to configure for optimal link performance. Recently, reinforcement learning has been proposed to automate link adaptation, where the transmission parameters are modeled as discrete arms of a multi-armed bandit. In this context, we propose a latent learning model for link adaptation that exploits the correlation between data transmission parameters. Further, motivated by the recent success of Thompson sampling for multi-armed bandit problems, we propose a latent Thompson sampling (LTS) algorithm that quickly learns the optimal parameters for a given channel state. We extend LTS to fading wireless channels through a tuning-free mechanism that automatically tracks the channel dynamics. In numerical evaluations with fading wireless channels, LTS improves the link throughout by up to $100\%$ compared to the state-of-the-art link adaptation algorithms.}

\end{abstract}


%
\IEEEpeerreviewmaketitle

\section{Introduction}

Link adaptation is a key feature of the wireless physical layer. In a fading wireless channel, link adaptation adapts the data transmission parameters, for example the modulation and coding scheme (MCS), to optimize the link performance in real time. Efficient and robust link adaptation is central to achieving the extremely high data rates supported by the state-of-the-art wireless networks~\cite{dahlman20185g}.

Wireless channels introduce stochastic impairments to the transmitted data symbols due to, for example, channel fading, phase rotation, and additive noise and interference. The wireless channel state is therefore a complex and time-varying property. Wireless systems estimate the instantaneous channel state from the outcome of previous transmissions as well as infrequent, measurement-based, channel reports. With link adaptation, these channel state estimates determine MCS used for encoding data packets in subsequent transmission intervals.

In the modern wireless networks, the physical layer also employs hybrid automatic repeat request (HARQ) for reliable communication. With HARQ, each received packet generates either an acknowledgement (HARQ ACK, or simply, ``ACK") or a negative ACK (NACK) indicating whether the receiver succeeded or failed in decoding the incoming packet. The chief purpose of the ACK/NACK feedback is to select the next set of data bits for transmission. Further, this feedback also provides a low-resolution indication of the underlying, or \emph{latent}, channel state. Several link adaptation schemes hence use the ACK/NACK feedback to refine their estimate of the channel state. In addition to the ACK/NACK feedback, the wireless transmitter may configure measurement-based channel feedback for a more fine-grained channel update. However, these reports can be expensive to generate and additionally consume precious wireless resources for signaling. Therefore, in this paper, we address link adaptation based only on the ACK/NACK feedback. We exemplify link adaptation in the context of cellular networks, although the general techniques are broadly applicable to other wireless protocols such as the IEEE 802.11 WiFi standard.

Link adaptation based on the ACK/NACK feedback falls under two categories: the classical outer loop link adaptation (OLLA) approach, and the recently-proposed reinforcement learning link adaptation (RLLA). With OLLA, the channel state is modeled in terms of an effective signal-to-interference-and-noise-ratio (SINR). The effective SINR is well-known to be a robust, low-dimensional, representation of the channel state~\cite{nanda1998frame}. OLLA hence maintains a dynamic, real-time, estimate of the SINR. For data transmission, OLLA maps the latest SINR estimate to the performance of each candidate MCS using a pre-generated offline link model (OLM)\footnote{Common OLLA implementations often short-circuit the OLM by pre-computing discrete \emph{switching thresholds}~\cite{nakamura2002adaptive,morales2009imperfect,saha2018link} . However, in this paper, we show that exploiting the full OLM output improves the link throughput.}~\cite{nakamura2002adaptive,morales2009imperfect,chung2001degrees,saha2018link}. In contrast to the channel state model adopted by OLLA, RLLA operates by learning the statistical behaviour for each MCS from the ACK/NACK feedback. Based on these statistical estimates, RLLA predicts the optimal MCS for data transmission in every transmission instance. In~\cite{combes2018optimal}, RLLA was rigorously modeled as a multi-armed bandit (MAB) optimization problem, which is the canonical framework for reinforcement learning\footnote{\review{Reinforcement learning is a vast field with numerous optimization techniques such as Q-learning, actor-critic algorithms, policy gradients, and others. These techniques learn a \emph{trajectory} of actions that optimize for a desired long-term objective. However, as shown in~\cite{combes2018optimal}, link adaptation is adequately modeled through the the elegant MAB framework, where the action in any step only influences the immediate reward. MABs are optimized using statistical techniques, the most prominent among which are upper confidence bound (UCB) or Thompson sampling based algorithms.}}.

The central shortcoming of OLLA is that it is challenging to tune  OLLA parameters for optimal performance. As an example, OLLA employs step-based SINR adjustments in response to the ACK/NACK feedback. The step sizes must be chosen carefully, since they affect how quickly OLLA converges to the optimal transmission configuration, and also the stability of this configuration. The most common step size selection mechanism uses a proxy configuration parameter known as the target block error rate (BLER), which itself is difficult to tune for good link performance~\cite{OPTIMAL_BLER}. In practice, the step size and BLER target parameters are tuned based on empirical evidence, which often leads to suboptimal link performance~\cite{OLLA_ANALYSIS,blanquez2016eolla}. 

RLLA mitigates some of the drawbacks associated with OLLA. In particular, RLLA automatically learns the statistical behavior of the candidate MCSs and thus reduces the dependence on tuning parameters. RLLA is hence more robust to arbitrary channel fading profiles than the previous techniques~\cite{combes2018optimal,saxena2019contextual,gupta2019link}. However, compared to the low-dimensional SINR model adopted by OLLA, RLLA uses a model that encodes each candidate MCS as a dicrete arm of a multi-armed bandit. The RLLA modeling complexity hence scales linearly with the number of candidate MCSs. This model is both resource-intensive and inefficient to train in real time. As a result, current RLLA algorithms can be slow to react to frequent channel variations. Additionally, current RLLA algorithms address channel fading through a sliding window protocol that needs to be tuned empirically for optimal performance.

\subsection{Contributions}

In this paper, we propose a new reinforcement learning algorithm for link adaptation. Our algorithm, latent Thompson sampling (LTS), extends the well-known Thompson sampling approach for online optimization. Compared to classical Thompson sampling, LTS encodes the environmnt through a low-dimensional latent state model. In the context of link adaptation, LTS adopts a probabilistic model of the channel SINR, whose parameters are learnt from the ACK/NACK feedback observed for previous transmissions. Since the SINR is a low-dimensional channel metric, LTS is able to quickly and reliably estimate the channel state from only a few ACK/NACKs. 
We make the following contributions in this paper:
\begin{itemize}
    \item We model the latent channel state, measured in terms of the instantaneous channel SINR, through a probabilistic model. LTS initializes this model based on the available channel information and subsequently refines it with the inferred channel response.
    \item We propose an efficient, OLM-based Bayesian update scheme to refine the SINR model with the decoding outcome obtained at the end of every transmission event.
    \item We develop upper bound on the throughput perfomance of LTS, where we show that the worst-case cumulative throughput loss scales with the square root of time.
    \item We extend the SINR estimation scheme to fading channels through a tuning-free mechanism. For this, we propose relaxing the SINR probability distribution in evry time step with an appropriately chosen smoothing function. The variance of this smoothing function is automatically configured using the channel Doppler estimates.
    \item We numerically evaluate the performance of LTS in terms of the average link throughput for frequency selective fading channels. Our results demonstrate that LTS significantly outperforms state-of-the-art OLLA and RLLA schemes for cellular networks.
\end{itemize}

\subsection{Organization}

The rest of this paper is structured as follows: In Section~\ref{sec:related_work}, we review the existing link adaptation schemes. Next, in Section~\ref{sec:model}, we describe the cellular link model and the link performance objectives considered in this paper. In Section~\ref{sec:thompson_sampling}, we introduce Thompson sampling in the context of link adaptation. Subsequently, in Section~\ref{sec:algorithm}, we introduce and describe our proposed LTS algorithm for link adaptation. We evaluate LTS numerically in Section~\ref{sec:numerical} and compare it with the state-of-the-art link adaptation algorithms. Finally, we conclude the paper in Section~\ref{sec:conclusion} and discuss future research directions.


\section{Related Work}
\label{sec:related_work}

Link adaptation has been an area of extensive research since the past two decades. Link adaptation has been studied in the context of both cellular networks and WiFi networks, although the terminology varies slightly across the two domains. Further, while cellular link adaptation generally favors measurement-based techniques, link adaptation schemes for WiFi tend to use sampling-based approaches. In the rest of this section, we review cellular link adaptation techniques, which are commonly referred to as OLLA, and the recently proposed RLLA schemes.

\subsection{Outer Loop Link Adaptation}

OLLA was first introduced in the context of third-generation cellular networks~\cite{nakamura2002adaptive}. Since then, OLLA has been adopted by fourth-generation (4G) and fifth-generation (5G) networks as well~\cite{pocovi2020channel}, and has also been proposed for satellite communication systems~\cite{tato2020link}. The effect of the target BLER parameter for OLLA on link throughput was studied analytically in~\cite{OPTIMAL_BLER} and~\cite{park2015optimizing} under a number of simplifying assumptions. At a high level, it was found that the optimal target BLER is a decreasing function of the SINR. While~\cite{OPTIMAL_BLER} argued for a broadly-applicable target BLER of $10\%$,~\cite{park2015optimizing} derived an analytical expression that maps a long-term average channel SINR to the optimal target BLER. However, obtaining optimal target BLERs with realistic channel impairments remains an open problem.

In addition to the difficulty of configuring the target BLER, OLLA also requires careful calibration of the step sizes. Large step sizes cause OLLA to oscillate around its steady state. On the other hand, small step sizes have better stability but slow the OLLA convergence towards a steady state. To address these issues,~\cite{duran2015self} proposed jump-starting the SINR estimate using an step size statistics from previous data flows. In~\cite{blanquez2016eolla}, an enhanced OLLA (eOLLA) scheme was proposed to dynamically adjust the step sizes based on the history of ACK/NACK feedback. The eOLLA scheme was shown to achieve better stability and modest throughput gains compared to classical OLLA. In~\cite{ohseki2016fast}, the step size was initialized with a large value that was then decayed exponentially with the time that a data flow was active. However, the channel variation profiles and data flow characteristics vary with time and across cellular deployments. As a result, it is challenging to configure OLLA step sizes that perform well in most scenarios.

Recently, reinforcement learning has also been proposed to automate OLLA configuration. In~\cite{wahls2013outer}, adaptive kernel regression was employed to automatically tune OLLA parameters for optimal link throughput. In~\cite{pulliyakode2017reinforcement}, reinforcement learning was employed to select from a set of optimal SINR adjustments modeled as independent arms of a multi-armed bandit. \review{Further,~\cite{saxena2019contextual,dong2018machine,khastoo2020neura,elwekeil2018deep,baldo2008learning,wang2010application} have proposed neural network models that learn the optimal data transmission parameters in real time.}

\subsection{Reinforcement Learning based Link Adaptation}

In contrast to reinforcement leaning-based OLLA enhancements, reinforcement learning has also been proposed for the statistical learning of MCS performance. These RLLA techniques rigorously model the MCSs as discrete arms of a MAB~\cite{combes2018optimal}. MABs elegantly encode decision-making in stochastic environments, where an agent should efficiently explore the available actions to maximize some environment-dependent reward function~\cite{slivkins2019introduction}. The key advantage with MAB-based link adaptation is that it directly optimizes for the desired link performance objective, for example the link throughput. These RLLA schemes hence do not rely on proxy tuning parameters such as the target BLER, nor on step-based heuristics. 

The first RLLA scheme was proposed in~\cite{combes2018optimal}, where an upper-confidence bound (UCB) scheme, graphical optimal rate sampling (G-ORS), was proposed. For efficient learning, G-ORS exploited the fact that the expected throughput as a function of the MCS exhibits a unimodal structure. Further,~\cite{combes2018optimal} provided theoretical performance bounds that are also asymptotically optimal in terms of convergence to the optimal MCS. Compared to UCB, Thompson sampling is a Bayesian MAB optimization technique that adopts probabilistic models of the unknown problem parameters. In empirical tests, Thompson sampling has been shown to outperform UCB-based algorithms~\cite{chapelle2011empirical}. Hence, several RLLA schemes build upon Thompson sampling for good link performance.~\cite{gupta2018low} proposed a Thompson sampling algorithm for link adaptation, where the correlation assumption between MCSs was dropped in favor of a simpler MAB model. However, neglecting the correlation between MCSs also hurts the learning rate of their proposed modified Thompson sampling (MTS) algorithm. Subsequently,~\cite{gupta2019link} proposed a constrained Thompson sampling algorithm that preserves the unimodality across the modeled MCS. In~\cite{paladino2017unimodal}, the G-ORS approach of~\cite{combes2018optimal} was translated to an analogous unimodal Thompson sampling (UTS) algorithm. While UTS has not been evaluated for link adaptation, it nevertheless outperforms G-ORS for several other scenarios. The problem of link optimization under constrained performance objectives was studied in~\cite{saxena2020bayesian, saxena2019constrained}, where suitable, analytically bounded, extensions of Thompson sampling were proposed. Further,~\cite{qi2019rate} proposed a Thompson sampling-based approach for link adaptation that compacts the search space for fast convergence. In this paper, we compare our proposed LTS approach with the MTS and UTS algorithms discusses above.

\review{The basic RLLA schemes optimize for a stationary environment, that is, where the channel response does not vary over time. However, in most practical deployments, the channel response is known to be nonstationary. To address these channel variations,~\cite{combes2018optimal} proposed a sliding window heuristic, which conditions the MCS selection only on the ACK/NACK feedback for the preceding few transmissions. The size of the sliding window is an algorithm tuning parameter that is chosen empirically. In contrast,~\cite{saxena2020bayesian} uses relatively frequent channel reporting to respond to channel fading. In this paper, we propose a Doppler-based mechanism for LTS that obviates the need for algorithm parameter tuning.}

\section{Model and Objectives}
\label{sec:model}

\review{We consider packetized data transmission over a wireless link}. In every transmission instance indexed by $t=1,2,\dots,T$, the wireless transmitter selects an MCS $m[t]\in\{1,\dots,M\}$. With MCS index $m[t]$, $D_{m[t]}$ bits are packed into a \emph{transport block} that is first encoded with a forward error-correcting code and bit-interleaved using a pseudorandom sequence to protect against stochastic noise and channel fading. The encoded and interleaved bits are mapped onto modulation symbols from a complex-valued alphabet of size $|F_{m[t]}|$ prescribed by the MCS. The sequence of modulated symbols is either truncated or zero-padded to completely fill the time-frequency resources allocated for transmission. The effective channel code rate is then given by 
\begin{align}
L_{m[t]} = \frac{1}{\Delta t\Delta f}\times\frac{D_{m[t]}}{\log_2|F_{m[t]}|} ,
\end{align}
where $\Delta t$ and $\Delta f$ denote the scheduled transmission duration and bandwidth respectively. The transmission symbols are subsequently multiplexed with known reference symbols for receiver-side channel compensation and loaded onto a discrete time-frequency grid of time-domain orthogonal frequency division multiplexing (OFDM) symbols and frequency-domain \emph{subcarriers}~\cite{dahlman20185g}. The \emph{data rate} for MCS $m$, i.e., the normalized amount of data carried, is given by
\begin{align}
r_{m[t]}\triangleq\frac{1}{\Delta t \Delta f}D_{m[t]}=L_{m[t]}\times\log_2|F_{m[t]}|.
\end{align}
The receiver attempts to decode the incoming signal and feeds back either a single-bit ACK, $c_{m[t]}[t]=1$, or a NACK, $c_{m[t]}[t]=0$, that signals a successful or a failed transport block reception respectively (typically determined at the receiver using a cyclic redundancy check (CRC) appended to the transport block). The ACK probability for a received transport block depends on the MCS $m[t]$ and the underlying channel state. Denoting the channel state for the $t^\text{th}$ transmission instance with $\theta[t]\in\Theta$, the ACK probability for MCS $m[t]\in[M]$ is given by
\begin{align}
    \mu\big(m[t],\,\theta[t]\big) = P\big[c_{m[t]}[t]=1\,\big|\,\review{m[t]},\theta[t]\big].
    \label{eq:ack_probability}
\end{align}

\subsection{Problem Objective}

For the transmission at time $t$, the \emph{realized} data throughput for a link with MCS $m[t]$ is
\begin{align}
    \mathcal{T}[t] &= r_{m[t]}\times c_{m[t]}[t],
\end{align}
that is, the normalized number of data bits delivered successfully to the receiver. We consider the link adaptation goal of maximizing the expected link throughput,
\begin{align}
    \text{maximize}\quad \mathbb{E}\big[r_{m[t]}\times c_{m[t]}[t]\big]
    \label{eq:goal}
\end{align}
To optimize the expected throughput, link adaptation predicts the optimal MCS at time $t$,
\begin{align}
    m^\star[t] &= \argmax_{m\in\{1,\dots,M\}}\,  r_{m}\times \widehat{\mu}_{m}[t] 
    \label{eq:MCS_selection}
\end{align}
where $\widehat{\mu}_{m}[t]$ is the predicted ACK probability of MCS $m$ at the transmitter.

\subsection{Effective SINR and OLM}

The ACK probability is \emph{a priori} unknown at the transmitter. With link adaptation, the transmitter needs to estimate the ACK probability associated with each MCS to select the optimal MCS that maximizes the expected throughput in every transmission instance. State-of-the-art OLLA approaches use the channel SINR as a robust \emph{latent} channel state metric to estimate ACK probabilities. In~\cite{nanda1998frame}, an \emph{effective} SINR metric (ESM) was proposed to compress the vector of per-subcarrier SINRs to a scalar value. Further, the ESM was shown to accurately parameterize the ACK probability for convolutional channel codes. Since then, several ESMs have been proposed that model a broad range of MCSs, where the compression parameters are learnt from a training dataset~\cite{brueninghaus2005link}. Recently,~\cite{saxena2018deep} also proposed an artificial neural network model that improves the ACK probability prediction compared to ESM-based techniques.

Cellular base stations and nodes maintain separate offline models that map an input ESM and MCS to an output ACK probability,
\begin{align}
    \mathcal{G}(m,\theta):(m,\theta)\mapsto \mu_m(\theta).
    \label{eq:olm}
\end{align}
Closed-form OLMs that takes the form of parameterized sigmoid-like functions have been proposed in~\cite{lembo2011modeling} and~\cite{carreras2018link}. Further, table-based OLM representations that discretize the feasible SINR range into a finite number of bins are commonly used in current cellular deployments~\cite{brueninghaus2005link}. Here, each bin corresponds to a corresponding ACK probability, which is typically stored in the form of a lookup table for fast access. The parameters for both the traditional table-based OLM and the analytical OLM are learnt by fitting the model to a training dataset either generated numerically or collected in the field.

\review{
\subsection{MAB Link Adaptation Model}
MABs are a special class of reinforcement learning models. MAB models the classical problem of picking the optimal action from a set of candidate actions, where executing an action in the environment generates a random-valued reward with an unknown distribution.  The choice of action in any round hence balances between exploring the actions with uncertain reward distribution, and exploiting the knowledge of rewards gained in the previous rounds. In contrast to the general reinforcement learning problem where the actions in any round influence future rewards, the MAB rewards are assumed to be independent and identically distributed across the rounds. The goal of the MAB agent is to maximize the cumulative reward over several rounds. The performance of any MAB optimization algorithm is then evaluated in terms of its \emph{regret}, which is the cumulative difference in the reward attained by the evaluated policy and the rewards for an oracle policy that picks the globally optimal action in every round. In the context of link adaptation, the transmitter serves as the MAB agent that operates in the wireless channel environment. The space of available MCSs constitute the candidate actions. In every round, which corresponds to a transmission interval, transmission using a selected MCS results in an ACK or a NACK, which is random variable parameterized by the unknown channel state. The link adaptation goal in~\eqref{eq:goal} is equivalent to maximizing the cumulative reward until time $T$. 
}

\review{
\section{Thompson Sampling for Link Adaptation}
\label{sec:thompson_sampling}

Thompson sampling is a Bayesian technique for MAB optimization, which was first proposed in 1933 in the context of clinical trials~\cite{thompson1933likelihood}. During the last decade, Thompson sampling has attracted renewed interest for optimizing MAB problems, for example in the context of recommender systems, online auctions, etc.~\cite{slivkins2019introduction}. This interest in Thompson sampling is also driven by the evidence of its state-of-the-art empirical performance~\cite{chapelle2011empirical}, and analytical breakthroughs related to its finite-time performance~\cite{kaufmann2012thompson, agrawal2013further}. Next, we introduce the general Thompson sampling principle, and summarize a couple of link adaptation algorithms based on this principle.

In the context of link adaptation, Thompson sampling models the ACK probability associated with each candidate MCS. These ACK probability estimates are used to predict the optimal, throughput-maximizing MCS. In every time step, Thompson sampling obtains sampled point estimates of the ACK probabilities. This sampling operation controls the \emph{exploration-exploitation tradeoff} of the MCS selection policy, by sometimes picking MCSs have uncertain reward performance. Over multiple transmission instances, Thompson sampling sequentially learns from the ACK/NACK feedback to refine the ACK probability model and select the optimal MCS.

\emph{MTS and UTS:} The MTS algorithm was proposed in~\cite{gupta2018low}, which employed the classical Thompson sampling algorithm for link adaptation. MTS uses the Beta distribution to model the ACK probability for each candidate MCS. Further, MTS reduces the modeling complexity by assuming that the MCSs are uncorrelated in terms of the observed ACK/NACK events. In contrast, the UTS algorithm was proposed in~\cite{paladino2017unimodal} based on the prior link adaptation formulation in~\cite{combes2018optimal}. The UTS algorithm exploits the correlation across the candidates MCSs by assuming that the throughput is a unimodal function of the MCS. In this manner, the UTS algorithm substantially reduces the algorithm complexity for link adaptation problems. In Section~\ref{sec:numerical}, we numerically compare the proposed LTS algorithm with MTS and UTS algorithms for link adaptation.

}

\section{Latent Thompson Sampling for Link Adaptation}
\label{sec:algorithm}

We propose a RLLA scheme, LTS, which models the channel SINR in terms of its probability distribution over a range of SINR values. In every time interval, LTS assigns a probability distribution to a range of feasible SINR values. Next, LTS obtains an SINR point estimate by sampling the SINR distribution and maps the point estimate to the ACK probabilities for each MCS using the OLM. The ACK probabilities are then used to predict the optimal MCS for transmission. The receiver signals a ACK/NACK feedback to signal the transmission outcome, which LTS uses to update the SINR probability distribution for the next time step. The SINR distribution iteratively concentrates around the true channel SINR, i.e., assigns higher probability density to the SINRs close to the true channel SINR. Further, LTS addresses channel fading by smoothing the SINR distribution to account for channel variations from one time interval to the next. The complete LTS algorithm is listed in Algorithm~\ref{alg:LTS}, and described subsequently.

\subsection{Probabilistic SINR Model}

The channel SINR is a real-valued quantity. Given that the channel is \emph{a priori} unknown at the transmitter, it is denoted by the SINR probability density function (PDF) $P_{\Theta[t]}[\theta]$ at the transmission time interval indexed by $t$. The initial SINR PDF, $P_{\Theta[0]}$, is uniformly distributed over a feasible SINR range if no channel information is available. On the other hand, if some knowledge of the channel SINR is available, for example from a recent CQI report, the initial SINR PDF $P_{\Theta[0]}$ also reflects this channel knowledge.

\subsection{SINR Point Estimates}

LTS calculates an estimated SINR in every transmission instance for predicting the optimal MCS. At the $t^\text{th}$ transmission instance, LTS generates the SINR sample
\begin{align}
    \widetilde{\theta}[t]\sim P_{\Theta[t]}.
\end{align}
In the initial few transmission intervals, the SINR sample has a relatively high variance owing to the limited (or absent) prior knowledge of the channel. The initial SINR samples hence explore the space of feasible SINRs. Over time, the SINR PDF is refined through Bayesian updates that are described in the next section. Subsequent draws are then more likely to be close to the peak of the SINR PDF, which concentrates around the true channel SINR.

Sampling from the SINR PDF is possible though several sampling techniques~\cite{cochran2007sampling}. Here we describe one such scheme, inverse transform sampling (ITS), which has commonly available and computationally efficiently implementations~\cite{olver2013fast}. ITS first calculates the the cumulative distribution function (CDF),
\begin{align}
  F_{\Theta[t]}\big[\theta\big] = \int_{x\leq\theta} P_{\Theta[t]}\big[x\big]dx.
 \end{align}
ITS then samples a uniformly distributed random variable, $u[t] \sim \mathcal{U}(0,1)$. Finally, ITS maps $u[t]$ to an SINR sample through the inverse SINR CDF, $\widetilde{\theta}[t] = F_{\Theta[t]}^{-1}[u[t]]$.

Sampling a SINR sample $\widetilde{\theta}[t]$ that underestimates the true channel SINR is likely to lower the instantaneous throughput. On the other hand, overestimating the channel SINR is even more detrimental to link performance: sampling a too-high SINR leads to aggressive MCS selection. Since most error rate curves have a sharp drop-off (also called waterfall curves~\cite{B_Wireless_Molisch}), aggressive MCS selection sharply reduces the instantaneous throughput. To mitigate this possibility, LTS adopts a~\emph{pessimistic} sampling approach, where the SINR estimate is chosen to be the minimum of the mean and sampled SINR values, i.e.,
\begin{align}
    \widehat{\theta}[t]=\min\{\,\widetilde{\theta}[t],\,E[\,\Theta[t]\,]\,\}.
    \label{eq:sinr_estimate}
\end{align}
Similar one-sided sampling approaches have earlier been studied in the context of optimistic Thompson sampling~\cite{may2012optimistic}.

\subsection{Optimal MCS Selection}

LTS uses the SINR point estimate $\widehat{\theta}[t]$ to predict the optimal MCS $m^\star[t]$ for the $t^\text{th}$ transmission interval. Similar to the OLLA MCS prediction step, LTS uses the OLM to map $\widehat{\theta}[t]$ to the MCS ACK probabilities $\widehat{\mu}_m[t]=\mathcal{G}^\text{Tx}\big(m,\widehat{\theta}[t]\big)\,\forall\,m\in[M]$. Consequently, to maximize the expected link throughput, the optimal MCS is obtained from~\eqref{eq:MCS_selection}. LTS can also easily address other link optimization goals, for example throughput maximization under a BLER constraint as considered in~\cite{saxena2020bayesian}. For these alternate performance goals, the appropriate suitable objective function is defined in place of~\eqref{eq:MCS_selection}. 

The transmitter processes the next set of data bits using $m^\star[t]$ and send them over the air. Subsequently, the receiver attempts to recover the data bits from the received signal processing and control signaling that indicates that MCS $m^\star[t]$ was used. The receiver then feeds back the binary ACK/NACK signal $c[t]\in\{0,1\}$ to the data transmitter.

\begin{algorithm}[t]
   \caption{LTS for Link Adaptation}
   \label{alg:LTS}
\begin{algorithmic}[1]
   \STATE {\textbf{Input:}}\, Data rates $r_m\,\forall\,m\in[M]$,\\
   $\quad\qquad$ Variance parameter $\sigma^2$.
   \STATE \textbf{Initialize:} SINR PDF $P_{\Theta[1]}$
   \FOR{ Time index $t=1$ {\bfseries to} $T$ }
   \STATE Determine an SINR estimate, $\widehat{
   \theta}[t]$, using~\eqref{eq:sinr_estimate}.
   \STATE Predict the optimal MCS, $m^\star[t]$, for $\widehat{\theta}[t]$ using~\eqref{eq:MCS_selection}.
   \STATE Transmit data with MCS $m^\star[t]$.
   \STATE Observe ACK/NACK feedback $c[t]$.
   \STATE Calculate the posterior SINR PDF, $P_{\Theta[t+1]}$.
   \ENDFOR
\end{algorithmic}
\end{algorithm}

\subsection{Posterior SINR Distribution}

LTS calculates the posterior SINR PDF using Bayesian updates to the prior SINR PDF. The posterior SINR PDF is $P_\Theta\big[\theta[t+1]\big]:=P_\Theta\big[\theta[t]\,\big|\,c[t]\big]$, i.e., the prior SINR PDF conditioned on the observed ACK/NACK feedback. Further, the likelihood of observing $c[t]$ is given by
\begin{equation}
\begin{aligned}
    P\big[c[t]\,\big|\,\theta\big]
    &=\begin{cases}
    P\big[c[t]=1\,\big|\,\theta\big],\,\,\,\,\qquad c[t]=1\\
    1-P\big[c[t]=1\,\big|\,\theta\big],\quad c[t]=0
    \end{cases}\\
    &=c[t]\times \mu(m^\star[t], \theta) + \big(1-c[t]\big)\times (1-\mu(m^\star[t], \theta)),
\end{aligned}
\label{eq:likelihood}
\end{equation}
where $\mu(m^\star[t], \theta)$ is defined in~\eqref{eq:ack_probability} as the ACK probability for MCS $m^\star[t]$ at SINR $\theta$. LTS estimates the ACK probability from the OLM defined in~\eqref{eq:olm} and available at the transmitter, i.e., $\widehat{\mu}_{m^\star[t]}[t] = \mathcal{G}^\text{Tx}(m^\star[t],\theta)$. The likelihood function can then be written as
\begin{align}
        P\big[c[t]\,\big|\,\theta[t]\big] \review{\approx}&      \,c[t]\times\mathcal{G}^\text{Tx}(m^\star[t],\theta) \nonumber\\
        &+ \big(1-c[t]\big)\times(1-\mathcal{G}^\text{Tx}(m^\star[t],\theta))
\end{align}

The posterior SINR PDF is then obtained from the Bayes' rule as
\begin{align}
     P_{\Theta[t+1]}&\big[\theta\big] = \frac{P\big[c[t]\,\big|\,\Theta[t]=\theta\big] \times P_{\Theta[t]}\big[\theta\big]}{P\big[c[t]\big]}.
     \label{eq:posterior_pdf}
\end{align}

The ACK probability $P\big[c[t]\big]$, is easily computed by marginalizing the corresponding $\theta$-conditional distribution, i.e., $P\big[c[t]\big]=\int_{\Theta[t]}P\big[c[t]\,\big|\,\theta\big]d\theta$. The posterior PDF can then be estimated through the self-normalized expression 
\begin{align}
    P_{\Theta[t+1]}\big[\theta\big] &=\frac{ P\big[c[t]\,\big|\,\Theta[t]=\theta\big] \times P_{\Theta[t]}\big[\theta\big]}{\int_{\Theta[t]}P\big[c[t]\,\big|\,\theta\big]P[\theta]d\theta}.
    \label{eq:posterior_pdf}
\end{align}

\review{
\subsection{Convergence Analysis}
}

\review{
In this section, we bound the expected link adaptation performance of the LTS algorithm. We characterize the finite-time regret of LTS, which measures the cumulative loss in throughput compared to an oracle policy that always selects the optimal MCS. We denote the true, unknown, channel SINR with $\theta^*$, which corresponds to the optimal MCS index $m^*=\argmax_{\{1,\dots,K\}} r_{m}\mu(m,\theta^*)$. Then, the finite-time regret for LTS is given by
\begin{align}
    \mathcal{R}(T;\theta^*) &= \mathbb{E}\bigg(\sum_{t=1}^T r_{m^*}\mu(m^*,\theta^*) - r_{m[t]}\mu(m[t],\theta^*) \bigg).
\end{align}
For Thompson sampling based optimization, a common regret analysis approach is to bound the \emph{Bayesian} regret, which is the expected regret over all parameter configurations, given by~\cite{slivkins2019introduction}
\begin{align}
    \mathcal{BR}(T) &= \mathbb{E}_{\theta^*\in\Theta}\big(\mathcal{R}(T;\theta^*)\big)
\end{align}
In the context of link adaptation, the Bayesian regret metric amounts to the expected LTS regret over the range of feasible channel SINRs. Hence, a bound on the Bayesian regret quantifies the maximum throughput loss experienced by an average wireless link.
\begin{theorem}
The Bayesian regret for LTS until time $T$ is upper bounded by  
\begin{align}
    \mathcal{BR}(T)\leq r_\text{max}\big(3M + \sqrt{6M\cdot T\log T}\big),
\end{align}
where $r_\text{max}=\text{max}\{r_1,\dots,r_K\}$. 
\label{th:regret}
\end{theorem}
}

\review{
Theorem~\ref{th:regret} shows that the Bayesian regret for LTS increases at most as fast as $O(\sqrt{T\log T})$, where $O(\cdot)$ is the big-Oh notation. Equivalently, the relative loss in throughput until time $T$ satisfies $O\big(\sqrt{\log T/T}\big)\xrightarrow{T\to\infty} 0$, and hence the algorithm is asymptotically optimal. In Section~\ref{sec:numerical}, we empirically evaluate LTS, where we demonstrate that this algorithm rapidly converges to the true channel SINR in only a few transmission time intervals. Next, we develop the proof for Theorem~\ref{th:regret}.
}

\review{
The convergence analysis for latent bandits is a topic of active research. The recent work in~\cite{hong2020latent} provides a timely update on this topic. The analysis in~\cite{hong2020latent} assumes that the reward distribution is sub-Gaussian, that is, the reward $R\sim P(\cdot|a,s)$ for action $a$ in latent state $s\in\mathcal{S}$ satisfies $\log \mathbb{E}\big( \text{exp}\big(\lambda(R-\mathbb{E}(R))\big)\big)\leq \lambda^2\sigma^2/2$, for all $\lambda\in\mathbb{R}$ and where $\sigma$ is the variance proxy. Under this sub-Gaussianity assumption, the Bayesian regret for the general latent Thompson sampling algorithm is shown to be upper bounded by
\setcounter{theorem}{0}
\begin{lemma}
\begin{align}
    \mathcal{BR}(T)\leq 3|\mathcal{S}| + 2\sigma\sqrt{6|\mathcal{S}|\cdot T\log T},
\end{align}
where $|\mathcal{S}|$ denotes the cardinality of $\mathcal{S}$.
\label{lem:bayes_regret}
\end{lemma}
}

\review{
The bound in Lemma~\ref{lem:bayes_regret} holds for discrete state spaces. Since the SINR is a real-valued quantity, this result is not directly applicable for link adaptation. Further, the LTS rewards correspond to the ACK/NACK feedback, which are Bernoulli-distributed in $\{0,1\}$. To bound the Bayesian regret for LTS in this context, we first show that the SINRs can be discretized without a loss in throghput performance. Next, we establish the sub-Gaussianity of ACK/NACK rewards and obtain its variance proxy. We use these two results to derive the stated bound.}

\review{
\setcounter{theorem}{0}
\begin{proposition}
There exists discretization $\Gamma$ for the range of feasible SINRs such that $|\Gamma|\leq M$ that uniquely encodes the the throughput-maximizing MCS for an arbitrary SINR.
\end{proposition}
\begin{proof}
  Consider the range of feasible SINRs $\Theta=[\theta_\text{min},\theta_\text{max}]$. We construct the discrete set of SINRs, $\Gamma=\{\gamma_1,\dots,\gamma_M\}$ through the following procedure: $\gamma_m=\{\theta:m^*(\theta)=\max_{m\in{1,\dots,M}}r_m\mu(m,\theta)\}$, that is, the set of SINRs for which MCS $m$ maximizes the link throughput. Ties are broken arbitrarily. Clearly, each element in $\Gamma$ corresponds to a unique, throughput-maximizing, MCS. Further, the cardinality of this discretization $|\Gamma|\leq M$, since there can be MCSs that do not maximize the link throughput for any SINR value. 
\end{proof}
As an example, the discretization $\Gamma$ can be constructed by querying the OLM with the range of feasible SINRs to obtain the corresponding expected throughputs for each MCS and constructing the discrete encoding above.
}

\review{
\setcounter{theorem}{1}
\begin{lemma}
The Bernoulli distributed random variable $X\sim \mathcal{B}(\mu)$ is sub-Gaussian with variance proxy $\sigma^2=1/4$.
\end{lemma}
\begin{proof}
From the definition of the Bernoulli distribution, we have that
\begin{align}
    \log \mathbb{E}(e^{\lambda(X-\mu)})&= \log \big(\mu e^{\lambda(1-\mu)}+(1-\mu) e^{\lambda(-\mu)}\big) \nonumber\\
    &= -\lambda\mu + \log(\mu e^\lambda + (1-\mu)).
    \label{eq:variance_proxy}
\end{align}
The expression in~\eqref{eq:variance_proxy} can be maximized with respect to $\mu$ to obtain
\begin{align}
    \max_{\mu}\, \log \mathbb{E}(e^{\lambda(X-\mu)})&=-1 + \frac{\lambda}{e^\lambda-1}-\log \frac{\lambda}{
    e^\lambda - 1} \nonumber \leq \frac{\lambda^2}{8},
\end{align}
where the last step uses Pinkser's inequality~\cite{cesa2006prediction}. Comparing with the definition, $X$ is sub-Gaussian with $\sigma^2=1/4$.
\end{proof}
}

\review{
Next, we define the following set of \emph{consistent} SINRs at time $t$:
\begin{align}
    C_t\leftarrow\bigg\{\gamma\in\Gamma:\mu_\gamma - \widehat{\mu}_\gamma(t) \leq \frac{1}{2}\sqrt{6 N_t(\gamma)\log t}\bigg\},
\end{align}
where $\Gamma$ is an arbitrary discretization of the SINR space, $\mu_\gamma$ is the expected reward for SINR $\gamma\in\Gamma$ obtained using the OLM, $\widehat{\mu}_\gamma(t)$ is the mean reward for all previous time steps where SINR $\gamma$ was selected, and $N_t(\gamma)$ is the number of times that $\gamma$ was selected until time $t$. The set $C_t$ contains the SINRs that contain the true channel SINR with a high probability. Further, we define an upper bound on the expected reward for each MCS, $U_t(m)=\argmax_{\gamma\in C_t}\mu(m,\gamma),\,m=1,\dots,M$. Following the approach in~\cite{russo2014learning, hong2020latent}, we decompose the regret at time step $t$ in the following manner,
\begin{align}
    &r_{m^*}\mu(m^*;\theta^*) - r_{m[t]}\mu(m[t];\theta^*) \nonumber \\
    &\leq r_\text{max}\big(\mu(m^*;\theta^*) - \mu(m[t];\theta^*)\big) \nonumber \\
    &= r_\text{max}\big(\mu(m^*;\theta^*) - U_t(m[t])+U_t(m[t]) - \mu(m[t];\theta^*)\big) \nonumber\\
    &\leq r_\text{max}\big(\big(\mu(m^*;\theta^*) - U_t(m*)\big)+\big(U_t(m[t]) - \mu(m[t];\theta^*) \big),
\end{align}
where $r_\text{max}=\max_{m\in\{1,\dots,M\}}r_m$ is the MCS corresponding to the largest throughput. This decomposition is of the same form as obtained in~\cite[Sec. 4.1]{hong2020latent}, and directly leads to the bound in Theorem~\ref{th:regret}. We omit a full description owing to space constraints, and refer the interested reader to~\cite{hong2020latent}.
}

\subsection{Extension to Fading Channels}

The posterior PDF encodes the knowledge of the channel SINRs at the end of the $t^\text{th}$ transmission round. For a stationary channel, the SINR does not vary over time. Successive ACK/NACK feedbacks for an arbitrary MCS are hence generated from a stationary ACK probability distribution induced by the channel SINR. However, for fading channels typically observed in practical wireless deployments, the channel SINR is non-stationary owing to the dynamics of propagation environment. When the propagation environment features sufficiently many propagation paths, these dynamics are closely approximated through independent Normal probability distributions on the real and imaginary signal components~\cite{B_Wireless_Molisch}. Further, the variance of the Normal distribution is proportional to the relative speed between the transmitter and the receiver~\cite{yucek2005doppler}. Here, we use this insight to allow probabilistic SINR tracking in fading channels.

LTS addresses SINR nonstationarity by relaxing the posterior SINR PDF through convolution with a Normal distributionin every time step. The updated SINR PDF is then given by
\begin{align}
      P^\text{(up)}_{\Theta[t+1]}= P_{\Theta[t+1]}\ast \mathcal{N}(0,\sigma^2),
      \label{eq:pdf_convolve}
\end{align}
where $\ast$ denotes the convolution operator and $\mathcal{N}(0,\sigma^2)$ is the Normal distribution with zero mean and variance $\sigma^2$. The operation in~\eqref{eq:pdf_convolve} corresponds to a priori modeling the channel fading as an autoregressive process with Gaussian innovation. The magnitude of SINR variations is a function of the relative speed between the transmitter and the receiver. Cellular deployments measure this relative speed in terms of the Doppler shift experienced by each link~\cite{hou2016radio,yucek2005doppler}. Hence, LTS automatically configures the variance parameter proportionally to the normalized Doppler estimate, where the proportionality factor needs to be chosen empirically.

\subsection{Algorithm Complexity}

\review{As discussed earlier, OLLA has a low implemenation complexity. In terms of memory, OLLA maintains a running estimate of the scalar SINR for each link. In every time interval, OLLA computes a new SINR estimate by adding (or subtracting) the step size from the previous SINR estimate based on the ACK (or NACK) feedback. Subsequently, in most OLLA implementations, the updated SINR estimate is mapped to the optimal MCS through a simple thresholding of the SINR space~\cite{nakamura2002adaptive}.}

\review{
In contrast to OLLA, the Thompson sampling based link adaptation schemes maintain a prior distribution over each individual MCS. This entails two distribution parameters for each MCS, that is, a total of $2K$ distribution parameters need to be stored in memory. In every time interval, $K$ Monte Carlo sampling steps are executed, one for each of the $K$ MCSs. The computational complexity of these sampling operations depends on several factors that have been analyzed in~\cite{anderson2018computational}. Subsequent to the sampling, the optimal MCS is computed by searching over the predicted throughputs for the candidate MCSs. The complexity of these algorithms is hence significantly higher than OLLA.}

\review{
Our proposed LTS algorithm needs to only maintain a single prior distribution, that is, over the scalar SINR metric. Further, in every time interval, LTS executes only a single Monte Carlo sampling step to obtain an SINR point estimate. This SINR is mapped to the ACK probabilities through a fast OLM-based table lookup. Compared to the existing RLLA schemes, LTS hence has a lower memory footprint (to store the SINR model) and a lower computational complexity for computing the optimal MCS.}

\subsection{Discussion}

LTS is an efficient link adaptation scheme that automatically adapts to the channel fading profile. Further, cellular LA with LTS fulfils the desired link adaptation characteristics of being fast, stable, and robust. The intuition behind fast convergence is easily developed from the \emph{waterfall} nature of error rate curves~\cite{B_Wireless_Molisch}: for a given MCS, the ACK probability equals one for sufficiently high SINRs and drops rapidly to zero as the SINR decreases. Hence, whenever an ACK is observed, LTS assigns zero probability densities to the subset of SINRs that correspond to a zero ACK probability for the selected MCS. Conversely, with a NACK, LTS assigns a zero probability density to the SINRs that always succeed for that MCS. Owing to the abrupt error rates predicted by the waterfall curves, a single ACK/NACK feedback can substantially narrow down the range of likely SINRs for fast convergence. Further, LTS has a stable steady-state performance. After convergence to the most likely SINR, subsequent ACK/NACK feedback signals only serve to concentrate the SINR PDF further. Hence, unlike OLLA, the SINR estimate with LTS does not fluctuate after convergence. Finally, LTS is robust to deployment- and hardware-specific impairments. Since LTS learns from the observed ACK/NACK feedback, it estimates the OLM-supported SINR that most closely matches the true channel state. Even in the presence of impairments, LTS accurately learns the true channel SINR as long as it is within the range of SINRs encoded by the OLM. 

\section{Numerical Results}
\label{sec:numerical}

In this section, we numerically evaluate the link adaptation scheme described in this paper. We simulate four link adaptation algorithms: (i) OLLA, which is state-of-the-art in cellular deployments, (ii) UTS, which exploits the unimodality of the throughput function~\cite{paladino2017unimodal}, ii) MTS, which employs Thompson sampling with uncorrelated arms~\cite{gupta2018low}, and (iv) LTS, our proposed  link adaptation algorithm that exploits the latent channel SINR state metric. We evaluate each link adaptation algorithm in terms of their expected throughput, $\mathcal{T}_N(t)$, for $t=1,2,\dots$, where the expectation is taken over $N=1000$ independent runs of each experiment.

The simulation code is written in Python and relies on the PY-ITPP library~\cite{py-itpp} for the communication and signal processing functionality. The PY-ITPP library also provides access to standardized open-source implementations of several small-scale fading stochastic channel models, two of which are considered here. We use Jupyter notebooks running on a remote server to execute the experiments and collect the results. Further, we use the Ray library~\cite{liang2017ray} to efficiently parallelize the experiments across several compute nodes running separate Linux kernels.

\subsection{Experimental Setup}

We evaluate each link adaptation algorithm in the following radio environments. First, we simulate a channel that is stationary in time and has a flat frequency response, also known as an additive white Gaussian noise (AWGN) channel. For this channel, we also compare the SINR evolution for the two SINR-tracking schemes, OLLA and LTS, respectively. Next, we simulate two frequency selective fading channels that respectively model pedestrian and vehicular radio environments. Finally, for the pedestrian environment, we also evaluate the effect of CQI reporting on the performance of the evaluated algorithms.

We consider downlink communication modeled on the LTE standard~\cite{TS36213}. The transport block sizes and MCS values are obtained from~\cite{TS36213} and use Turbo codes for encoding the data bits. For the frequency selective and time-varying SINR experiments, we simulate two different channel models: a pedestrian environment where the relative speed between the transmitter and the receiver is $3$ km/h, using the ITU\_PEDESTRIAN\_A channel model~\cite{rec1997itu}, and a vehicular environment with a relative speed of $30$ km/h modeled using the ITU\_VEHICULAR\_B channel model~\cite{rec1997itu}. The pedestrian and vehicular environments respectively emulate low frequency selectivity with gradual fading, and high frequency selectivity with rapid fading.  The complete simulation parameters are listed in Table~\ref{tab:sim_par}.

We make the following assumptions throughout the numerical evaluations. We assume perfect synchronization between the transmitter and the receiver. Further, we assume full-buffer traffic, that is, the transmitter has sufficient data bits in the buffer to completely fill the choice of transport block in every transmission instance. We assume that the ACK/NACK feedback is available at the transmitter without any signaling delay. We assume that the channel is known perfectly at the receiver, both for decoding the data bits and for estimating the CQI. We assume perfect control channel feedback so that the CQI and the ACK/NACK feedback is always signaled successfully to the transmitter. For CQI generation, we ignore any systematic bias between the OLM employed by the transmitter and the receiver. Further, we only consider the first ACK/NACK transmission and ignore any retransmissions. For fading channels, we assumes a block fading profile, i.e., the channel response stays constant within a transmission interval but varies across the intervals. Further, we assume that the transmitter has perfect knowledge of the normalized channel Doppler, $\gamma=\frac{f_c\cdot \Delta v}{c}\cdot T_s$, where $f_c$ is the carrier frequency, $\Delta v$ is the relative speed of the receiver with respect to the transmitter, and $T_s$ is the sampling interval. 

\subsection{Algorithms}

The OLLA parameters are configured based on the previous literature. In particular, The target BLER for OLLA is chosen to the commonly used value of $0.1$~\cite{OLLA_ANALYSIS,duran2015self,blanquez2016eolla}. For step size configuration, different studies have used values ranging from $0.01$ dB uup to $1.0$ dB. As discussed in previous sections, the OLLA step size strikes a balance between the convergence rate and the stability of the algorithm. We experimented with several different values for OLLA, and found that a value of $\Delta=0.5$ to be suitable for link throughput optimization. Finally, the OLLA SINR estimate is constrained within maximum and minimum SINR values obtained from the range of SINRs supported by the OLM. These values effectively constitutes the feasible SINR region for the experiments.

MTS and UTS do not require any additional configuration for time-stationary channels. However, with fading channels, these algorithms adapts to the channel dynamics by employing a moving window over the transmissions and their corresponding ACK/NACK outcomes~\cite{paladino2017unimodal}. A moving window retains the historical MCSs and observed ACK/NACK outcomes for the previous few time steps and discards the rest. This historical data is used to predict the optimal MCS for the current time step. The size of the moving window, $L$, varies inversely with the channel coherence. However, an exact expression for $L$ is not available and heuristics are applied to choose an appropriate value. We select $L$ by experimenting with several possible values with each fading channel, and pick those values that provide the best link performance.

LTS uses a probabilistic SINR model. We implement LTS through a discretization of the SINR probability distribution. This implementation approach has several advantages: first, a discretization helps overcome the challenge of expressing the SINR PDF in a closed form. Second, the cellular standards limit the available MCSs to a finite set of discrete MCS values. A sufficiently fine SINR discretization hence does not suffer any performance loss compared to the SINR PDF. Finally, a discrete SINR probability mass function (PMF) can be used with computationally efficient sampling schemes. We discretize the feasible SINR range into $P$ SINR bins. The SINR PMF at time interval $t$ is then denoted by
\begin{align}
  P_{\Theta_k}[\theta] := P_{\Theta_k}[{\theta}\in\Theta_p]\quad\forall\,p\in[P],
\end{align}
where $\Theta_k$ is the set of SINRs modeled by the $p^\text{th}$ bin and $\cup_{p\in[K]}\Theta_p$ is the range of feasible SINRs.

As discussed in Section~\ref{sec:algorithm}, we use ITS to obtain a point SINR estimate from the SINR PMF~\cite{cochran2007sampling}. To calculate the posterior SINR PMF, the integration in~\eqref{eq:posterior_pdf} is replaced by a summation, and the Normal distribution employed by~\eqref{eq:pdf_convolve} is suitably discretized for convolving with the posterior PMF.

\review{We configure the variance parameter for LTS by linearly scaling the normalized Doppler with a constant scaling factor of $10^4$. This scaling factor was obtained empirically in the context of the pedestrian channel. For this, we simulated various LTS variance configurations and picked the variance that maximized the link throughput in the pedestrian scenario. Subsequently, while configuring LTS for the vehicular channel, we computed a new LTS variance by simply scaling the vehicular Doppler with the same factor of $10^4$, that is, without any additional simulations. We observed that with this configuration approach, LTS retains good performance for the vehicular channel as well. An intuitive reason for this is that the channel dynamics are indeed a factor of the relative Dopplers for the respective links. With this approach, the scaling factor needs to be obtained only once, and can then be used to automatically tune LTS configuration for several operating environments that have different Dopplers.}

\begin{table}[t]
\caption{Simulation Parameters}
\vspace{-5mm}
\label{tab:sim_par}
\centering
\begin{tabular}[t]{p{4cm}p{3.5cm}}
\toprule
Scenario Parameter & Value\\
\midrule
Carrier Frequency, $f_c$                      & $2$ GHz\\
Number of Subcarriers             & $72$\\
FFT Size, $M$                 & $128$ \\
Subcarrier Spacing, $\Delta f$              & $15$ kHz\\
Subframe Duration, $\Delta t$ & $1$ ms\\
Feasible SINR Range                  & [$-10$ dB, $20$ dB] \\
Channel Models                & AWGN, ITU\_PEDESTRIAN\_A~\cite{rec1997itu}, ITU\_VEHICULAR\_B~\cite{rec1997itu}\\
Relative Speeds $\Delta v$                       & $3$ km/h (pedestrian), $30$ km/h (vehicular) \\
\toprule
Configuration Parameters & Value\\
\midrule
OLLA Target BLER, $\eta$     & $0.1$ \\
OLLA Step Size Parameter, $\Delta$      & $0.5$ dB\\
OLLA Minimum SINR    & $-8.5$ dB \\
OLLA Maximum SINR    & $18.0$ dB \\
MTS/UTS Window Size, $L$       & $\{30,100\}$\\
LTS SINR bin size, $P$ & $1.0$ dB \\
\bottomrule
\end{tabular}
\end{table}

\subsection{AWGN Channel}

We first consider an AWGN wireless channel, where the SINR does not vary with time and where the channel response is flat across the frequency. The true channel SINR is configured to be $10$ dB. We assume that the true channel SINR is unknown at the start of the experiment and that no CQI report is available. The goal of link adaptation is therefore to maximize the link throughput by quickly converging to the true SINR. The initial state of algorithms is configured in the following manner: for each experiment, the initial OLLA SINR is selected randomly from the feasible SINR range. For the UTS, MTS, and LTS algorithms, the initial probability distributions for the modeled parameters are uniform over the respective parameter range.

\begin{figure}[t!]
\centering
\subfigure[Link throughput $\mathcal{R}_N(t)$.]{
    \includegraphics[width=\columnwidth]{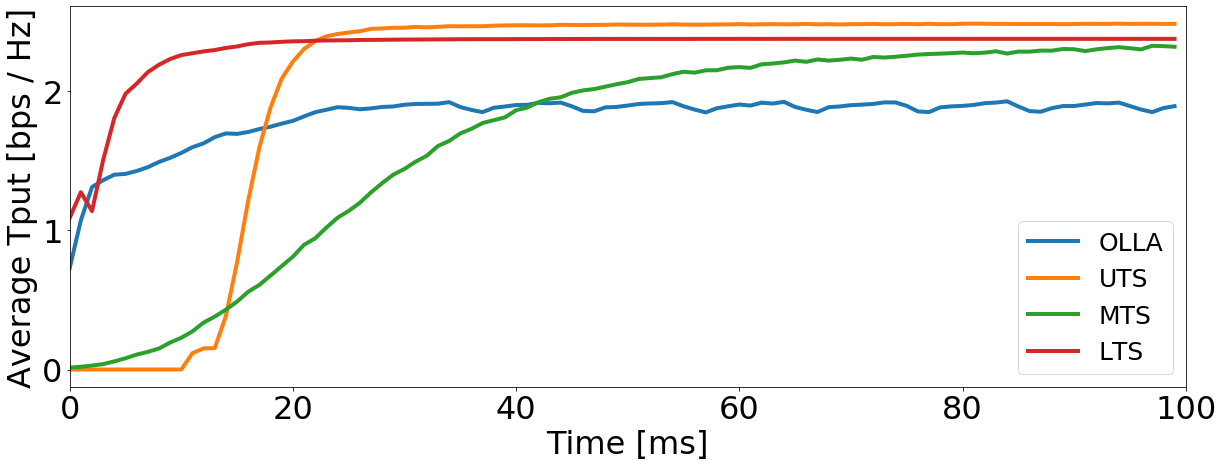}
    \label{fig:awgn_channel_tput}
}
\subfigure[Estimated channel SINR, mean values (solid curve) and one standard deviation (shaded region).]{
    \includegraphics[width=\columnwidth]{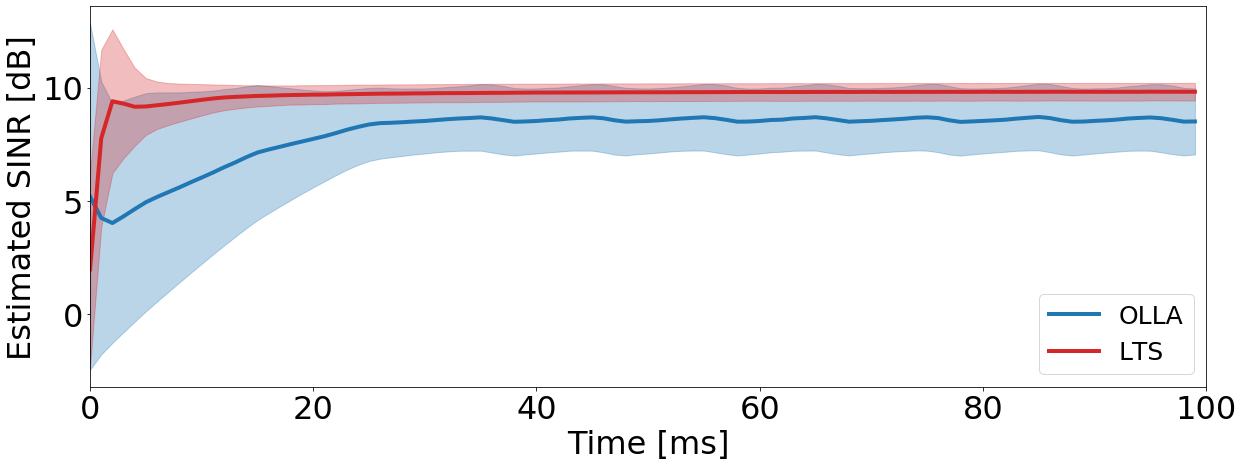}
    \label{fig:awgn_channel_sinr}
}
\label{fig:awgn_channel}
\caption{Link throughput (top figure) and estimated SINR (bottom figure) averaged over $N$ independent runs of an experiment with the AWGN channel. LTS converges quickly to the optimal throughput and has a stable steady state performance. In terms of SINR evolution, the SINR estimate for LTS concentrates around the true channel SINR and has a smaller steady-state variance than OLLA.}
\end{figure}

The realized throughput for each link adaptation scheme is illustrated in Fig.~\ref{fig:awgn_channel_tput}, where the throughput in each step is averaged over $N$ independent experiments. The average throughput for OLLA (solid blue curve) increases from the initial value and settles around a stable-state value after a few tens of transmissions. Compared to OLLA, UTS (solid orange curve) has poor performance for approximately the first $20$ time intervals, where the UTS link throughput is close to zero. The reason for this behaviour is that UTS initially assigns identical ACK probability distributions to each MCS. UTS hence predicts high expected throughput for the more aggressive MCSs. These MCSs inevitably fail owing to the given channel state, which results in a zero throughput. Subsequently, UTS probes each MCS sequentially until it obtains the optimal throughput-maximizing SINR after $~20$ time intervals. UTS subsequently maintains the optimal link throughput for rest of the time intervals. The MTS algorithm (solid green curve) achieves an overall poorer performance than UTS, but outperforms OLLA in the latter half of the experiment. The proposed LTS (solid red curve) outperforms each of the competing techniques in terms of the link throughput. LTS converges quickly to the optimal throughput and maintains a stable performance subsquently. For the later time intervals, LTS is observed to achieve a slightly smaller throughput than UTS. This is because LTS performance is fundamentally constrained by the discretization of the OLM lookup table, which quantizes the MCS ACK probabilities. The OLM model can be improved to overcome this performance gap, for example by using data-driven OLMs~\cite{saxena2018deep}.

The OLLA and LTS algorithms explicitly estimate the channel SINR. In Fig.~\ref{fig:awgn_channel_sinr}, we plot the SINR evolution for these algorithms. Solid lines denote the mean SINR across the experimental runs and the dotted lines denote a single standard deviation distance from the mean. Here, OLLA is observed to gradually converge towards the true channel SINR and simultaneously reduce the SINR estimation variance. In the case of LTS (red curves), the SINR estimate is highly uncertain at the beginning of the experiment. However, the SINR estimate quickly concentrates around the true channel SINR. 

\subsection{Frequency Selective Fading Channels}

Next, we discuss the numerical results for the pedestrian and vehicular channel models. For the OLLA algorithm, the step size parameter is kept at $\Delta=0.5$ dB. The window sizes for UTS and MTS are chosen to be $L=100$ and $L=30$ respectively for the pedestrian and vehicular channels. The LTS variance parameter is configured to be $\sigma=0.3$ and $\sigma=3.0$ for these two fading channels based on the discussion earlier.

\begin{figure}[t]
\centering
\includegraphics[width=\columnwidth]{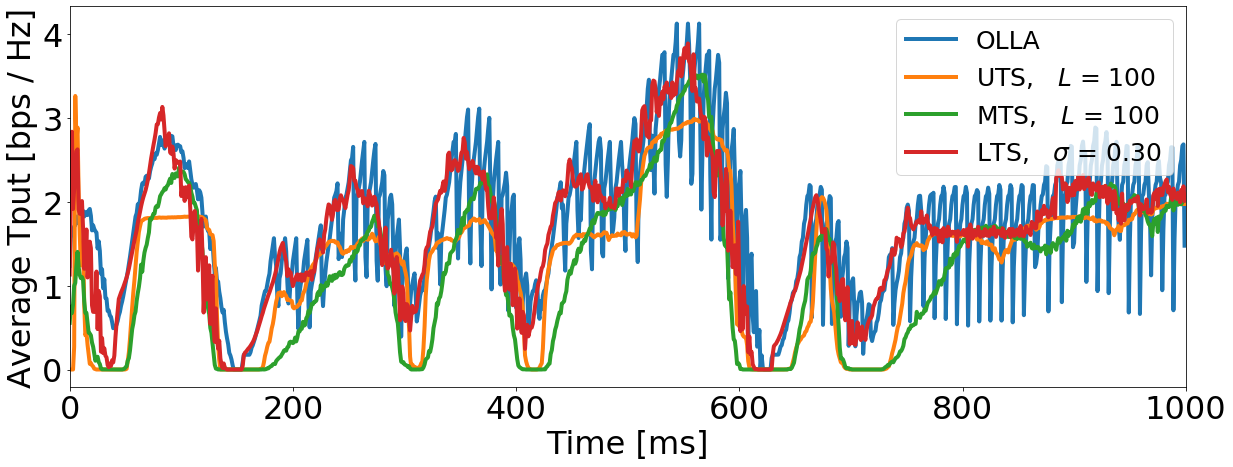}
\caption{Link throughput averaged for pedestrian channel with a relative speed of $3$ km/h. Compared to OLLA, UTS and MTS, LTS adapts faster to the channel variations and is stable across the time intervals.}
\label{fig:freqsel_fading_pedestrian}
\end{figure}

In Fig.~\ref{fig:freqsel_fading_pedestrian}, the average realized throughput for OLLA, UTS, MTS, and LTS algorithms are illustrated for the pedestrian channel with blue, orange, green, and red curves respectively. We observe that OLLA achieves good throughput performance for the initial few time intervals, but exhibits large fluctuations in later intervals. In contrast, UTS has a relatively more stable throughput performance. However, the peak throughput achieved by UTS is up to $30\%$ lower than OLLA. Compared to both OLLA and UTS, MTS achieves a lower link throughput. In contrast, LTS demonstrates the best performance among the evaluted link adaptation algorithms. The throughput with OLLA is both higher then competing algorithms, and stable across the time intervals.

\begin{figure}[t!]
\centering
\includegraphics[width=\columnwidth]{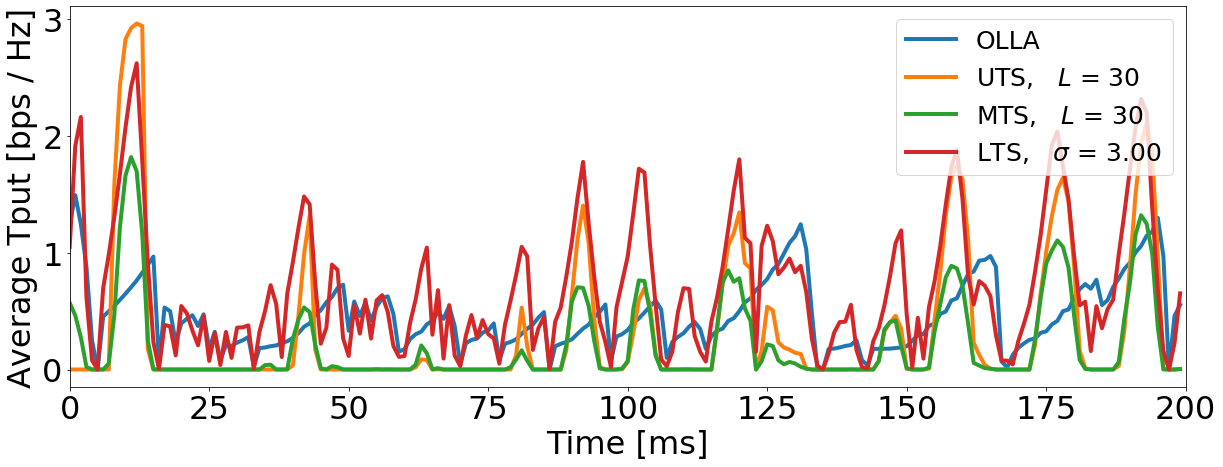}
\caption{Link throughput averaged for vehicular channel with a relative speed of $30$ km/h. Compared to OLLA, UTS, and MTS, LTS increases the average link throughput by up to $100\%$.}
\label{fig:freqsel_fading_vehicular}
\end{figure}

We illustrate the link adaptation performance for the vehicular channel in Fig.~\ref{fig:freqsel_fading_vehicular}. Here, the average throughput realized by LTS is up to $100\%$ higher than that achieved by the competing algorithms. Further, MTS is observed to have a zero throughput for a large number of time intervals since it does not adapt fast enough to the channel. As a result, MTS underperforms OLLA for most of the transmission duration.

\subsection{Explicit Channel Reporting}

\begin{figure}[t]
\centering
    \includegraphics[width=\columnwidth]{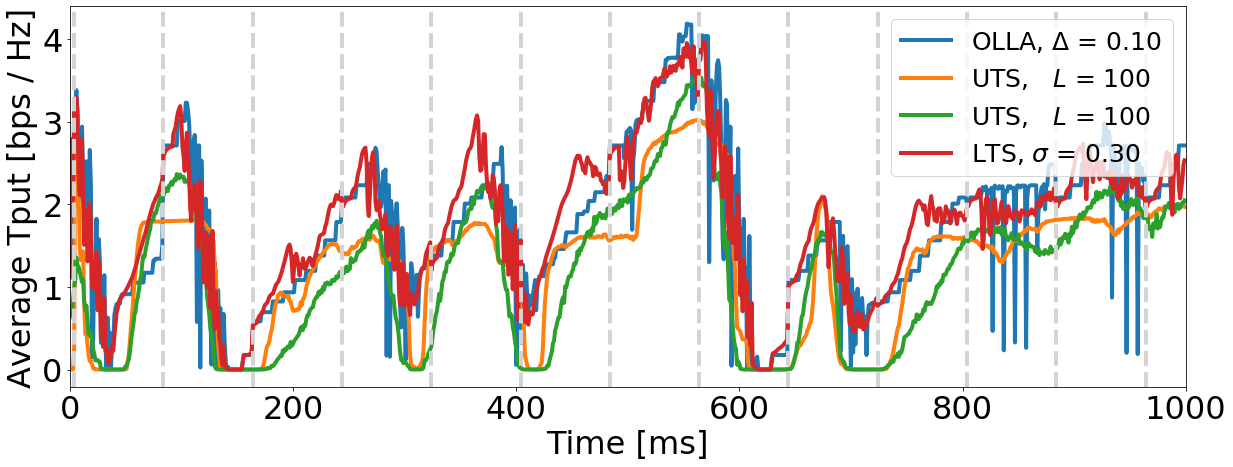}
\caption{Link throughput averaged for pedestrian channel when CQI reports are signaled every $80$ ms with a signaling delay of $4$ ms (indicated by dashed gray lines). At the cost of higher signaling overhead, CQI reports can improve OLLA performance to match the LTS performance.}
\label{fig:pedestrian_cqi}
\end{figure}

We consider the effect of CQI reports on the link adaptation performance for the pedestrian channel. We assume that the CQI reports are available at the transmitter every $80$ ms, with a CQI signaling delay of $4$ ms after the channel measurement. The CQI reporting instances are indicated with dashed gray lines in Fig.~\ref{fig:pedestrian_cqi}. Here, we choose a smaller step size $\Delta=0.1$ dB to OLLA. This is because with periodic CQI reports, OLLA can quickly adapt to larger channel variations. Further, smaller step size allows stable throughput performance. The UTS and MTS algorithms do not have a mechanism to incorporate CQI reports, and are illustrated in the figure for completeness. The performance results for OLLA, UTS, MTS, and LTS are illustrated in Fig.~\ref{fig:pedestrian_cqi} with blue, orange, green, and red curves respectively. We observe that OLLA achieves a good throughput performance by relying on CQI. Compared to Fig.~\ref{fig:freqsel_fading_pedestrian}, OLLA quickly ramps up the SINR estimate when the CQI reports are available, for example at $t=80, 240, 640$ ms. OLLA thus overcomes its performance gap compared to LTS with the additional channel knowledge provided by the CQI reports. In contrast, the impact of CQI reports on LTS is rather limited. For a few time intervals (for example, $t=480, 800$ ms, the LTS throughput is slightly helped by the availability of CQI reports. However, these gains are insignificant compared to the signaling overhead required to carry the CQI reports. LTS can thus mitigate the need for CQI reports for this scenario and achieve near-optimal throughput based only on the ACK/NACK feedback.

\section{Conclusions and Future Work}
\label{sec:conclusion}

We have proposed LTS, a new approach for link adaptation in cellular systems. LTS models an SINR probability distribution to estimate the latent channel SINR based on the ACK/NACK feedback. By optimally exploiting an OLM at the transmitter, the SINR distribution concentrates quickly around the true effective channel state. Further, LTS can be made to adapt to a fading channel by relaxing the SINR PDF proportionally to the channel Doppler. Compared to state-of-the-art OLLA and RLLA techniques, LTS (i) does not require any configuration tuning, (ii) converges faster to the optimal SINR estimate, and (iii) demonstrates a stable steady-state performance. Through numerical evaluations, we demonstrate that LTS significantly outperforms the existing link adaptation algorithms in terms of the average link throughput.

\review{Further work on LTS can involve the selection of an appropriate SINR PDF update function for diverse channel fading profiles. For example, with a dominant line-of-sight component, the assumption of Gaussian innovation for channel fading might not be the optimal choice. In that case, another suitable function could be applied for the posterior SINR PDF update. Similarly, mork work is required to address abrupt channel variations owing to large-scale obstructions and interference signals. Finally, the numerical evaluation for LTS could be carried out in the presence of several other impairments observed in practice, such as signaling delays, OLM biases, channel estimation errors, interference, and others.}

The latent modeling approach adopted by LTS can be applied to other domains, and to other problems in wireless communications. For example, with beam-based transmission commonly used in fifth generation (5G) cellular systems, the beam space may described by a compact angular subspace. Then, the relative performance of a beam may be learnt efficiently through a similar Bayesian scheme. Similar approaches could also be proposed in the context of frequency sub-band selection, rank adaptation, and spatial multiplexing for multi-user MIMO.


%





\ifCLASSOPTIONcaptionsoff
  \newpage
\fi



%
\newpage

\bibliographystyle{IEEEtran}

\bibliography{refs}

\newpage




%








\end{document}